\documentclass[aps,prd,twocolumn,nofootinbib,amsmath,amssymb]{revtex4-1}
\usepackage{CJK}                     % this comes first
\usepackage[dvips]{graphicx}         % this comes second
\usepackage{mathptmx}                % math+times fonts
\usepackage{bm}
\usepackage{physics}
\usepackage{dcolumn}                 % align table columns on decimal point
\usepackage{multirow}                % tables
\usepackage{xcolor}
\usepackage{hyperref}
\hypersetup{
breaklinks=true,colorlinks=true,
linkcolor=blue,citecolor=blue,filecolor=magenta,urlcolor=black}

\def\be{\begin{equation}} \def\ee{\end{equation}}
\def\bal#1\eal{\begin{align}#1\end{align}}
 \def\xv{\bm{x}} 
\def\et{\mathbf{e}}

\def\eps{\varepsilon}
\def\phi{\varphi}
\def\la{\lambda}
\def\al{\alpha}
\def\om{\omega}
\def\ms{\,M_\odot}
\def\mmax{M_\text{max}}

\def\km{\,\text{km}}
\def\fm3{\,\text{fm}^{-3}}
\def\gc3{\,\text{g/cm}^3}
\def\mfm{\,\text{MeV}\,\text{fm}^{-3}}
\def\taugw{\tau_\text{GW}}
\def\deps{\delta\epsilon}
\def\xp{x_p}

\begin{document}

\title{Non-radial oscillations and gravitational wave emission
of hybrid neutron stars}

\begin{CJK*}{UTF8}{gbsn}

\author{Zi-Yue Zheng (郑子岳)}%\email{Email:ziyuezheng@cug.edu.cn}
\author{Ting-Ting Sun (孙婷婷)}
\author{Huan Chen (陈欢)}\email{Email:huanchen@cug.edu.cn}
\author{Jin-Biao Wei (魏金标)}
\affiliation{
School of Mathematics and Physics, China University of Geosciences,
Lumo Road 388, 430074 Wuhan, China}

\author{G. F. Burgio}
\author{H.-J. Schulze}
\affiliation{
INFN Sezione di Catania, Dipartimento di Fisica,
Universit\'a di Catania, Via Santa Sofia 64, 95123 Catania, Italy
}

\begin{abstract}
We investigate non-radial oscillations of pure and hybrid neutron stars,
employing equations of state of nuclear matter
from Brueckner-Hartree-Fock theory,
and of quark matter from the Dyson-Schwinger quark model,
performing a Gibbs construction for the mixed phase in hybrid stars.
Characteristic differences between neutron-star and hybrid-star
$g_1$-mode oscillation frequencies, damping times, and gravitational wave strains 
are pointed out. Prospects of observations are also discussed.
\end{abstract}

%\bigskip
%\noindent Key-words:
% Brueckner-Hartree-Fock(BHF) theory,
% Dyson-Schwinger equation,
% Radial oscillation
%\bigskip

%\noindent PACS Number(s): 14.20.Dh, 14.40.Aq, 11.10.Lm, 12.39.Fe

\maketitle
\end{CJK*}

%------------------------------------------------------------------------------
\section{Introduction}

The interior of a neutron star (NS) can reach several times
the nuclear saturation density $\rho_0 \simeq 0.16\fm3$.
Therefore, there might exist a phase transition to deconfined quark matter (QM)
in the NS core \cite{Annala20}.
In general, once the equation of state (EOS) of nuclear matter (NM) is known,
one can compute the structure and properties of NSs.
Unfortunately, so far,
due to the lack of exact computation
dealing with the nonperturbative strong interaction,
the EOS of high-density NM remains an open theoretical problem.
There are many theoretical models for the NS EOS
that can meet the observational and experimental constraints,
see, e.g., \cite{Burgio21} %\cite{2018Chap6}
for a recent review.
For NM in the hadron phase,
popular EOSs are based on
relativistic mean field models \cite{Ring96},
phenomenological models based on energy-density functional theory
with generalized Skyrme effective forces \cite{Potekhin13},
Brueckner-Hartree-Fock (BHF) theory
\cite{Li08b,Kohno13,Fukukawa15,Lu19,Liu22},
the variational method (APR) \cite{Akmal98},
the self-consistent Green's functions approach
\cite{Carbone13},
and chiral effective field theory %($\chi$EFT)
\cite{Hebeler10,Coraggio14,Wellenhofer14,Drischler15}.
For QM, EOSs are mainly obtained with
the MIT bag model \cite{Chodos74},
the Nambu-Jona-Lasinio model \cite{Buballa05,Klahn13,Klahn15},
the perturbative QCD \cite{Kurkela09,Fraga14,Jimenez19},
and the Dyson-Schwinger equations (DSEs)
\cite{Roberts94,Alkofer01,Chen11,Chen12,Chen15}.

To test the various theoretical models of NSs and hybrid stars (HSs),
we have to resort to observations.
The mass-radius relation is one of the most straightforward and
simple observables,
which can be theoretically obtained by solving the Tolman-Oppenheimer-Volkov
(TOV) equations combined with the EOS.
Recently, several observed pulsars with masses above two solar masses
\cite{Demorest10,Antoniadis13,Fonseca16,Cromartie19}
put firm lower limits on the maximum mass of NSs.
Some theoretical analyses of the NS merger event GW170817
even deduce an upper limit on the maximum mass of about 2.2--2.3$\,\ms$
\cite{Shibata17,Margalit17,Rezzolla18,Shibata19,Shao20},
albeit with large theoretical uncertainty.
In 2019, new constraints on the radius were provided by the NICER
(Neutron Star Interior Composition Explorer) mission,
which reported Bayesian parameter estimations of the mass and equatorial radius
of the millisecond pulsar PSR J0030+0451 \cite{Riley19,Miller19},
and recently on PSR J0740+6620 with mass $2.08_{-0.07}^{+0.07}\,\ms$
\cite{Cromartie19,Fonseca21,Miller21,Riley21,Pang21,Raaijmakers21}.

However, it is hard to distinguish HSs and pure NSs
from the mass-radius relations,
since theoretically the differences between them are small,
and even masked by the uncertainties of pure NSs with various models
\cite{Alford09}.
Therefore, one needs other observables to reveal the interior of NSs.

NSs are sources of electromagnetic waves in all wavelengths,
and also emitters of both continuous and inspiral gravitational waves (GWs).
In particular, when NSs experience violent processes such as accretions,
radial and non-radial oscillations, glitches, and even NS mergers
\cite{Abbott17a,Abbott17b},
they are expected to emit strong enough signals that can be observed.
%and might reveal the structure and properties of NSs.
After the first direct observation of GWs from a binary black hole (BH) merger
\cite{Abbott16},
more and more GW signals were detected,
including NS mergers
\cite{Abbott17a,Abbott17b,Abbott19,Abbott21},
which has opened a new window on NS observation
by using GWs as probes of their internal structure.

In our previous work \cite{Sun21} we investigated radial oscillations,
and found a clear difference of their frequencies between pure NSs and HSs.
The radial oscillation of NSs is the simplest oscillation mode
without direct GW radiation,
but might couple with and amplify GWs \cite{Passamonti05,Passamonti07}
and modulate the short gamma ray burst (SGRB) \cite{Chirenti19}.

On the contrary,
the non-radial oscillation (NRO) of a star can directly produce GW signals
\cite{Thorne67},
not only in NS mergers or supernova explosions.
Therefore, the study of NROs of isolated NSs
may provide us with direct and stable observations to understand
the structure and properties of high-density NM,
the strong interaction, and GWs.
In this work we further study the NROs of NSs,
specifically the quadrupole oscillations
which are stable sources of GW radiation.

The theoretical study of NROs in general relativity
was first proposed by Thorne \cite{Thorne67,Price69}.
After that,
a series of rigorous mathematical analyses were carried out
by Detweiler and Lindblom \cite{Lindblom83,Detweiler85},
whereas Chandrasekhar and Ferrari proposed a simpler calculation
and obtained some properties of NROs of compact stars \cite{Chandrasekhar91}.
In recent years,
many investigations of NS NROs were carried out,
for example
\cite{Burgio11,Flores17,Jaikumar21,Zhan21,Constantinou21,Zhao22,Zhao22b,
Kunjipurayil22}.
They show that some eigenfrequencies of the NROs of NSs
are within the sensitive range of some current detectors.
Theoretically, for a non-rotating NS,
the eigenmodes of oscillation are divided into
$g$ mode, $f$ mode, and $p$ mode,
which indicate the various dominant restoring forces for the perturbations.
The $g$-mode eigenfrequencies are relatively small,
$\sim{\cal O}(1~\text{kHz})$,
and provide us with an appropriate observable.
The buoyancy acts as the $g$-mode restoring force
to bring disturbed fluid elements back into equilibrium,
and generally its frequency depends on the particle fraction gradient
and temperature \cite{Reisenegger92}.
This effect is more intense in HSs than in pure NSs \cite{Jaikumar21,Kumar23}.
Therefore, we can expect different characteristics of the $g$ modes
between NSs and HSs,
and we will mainly focus on this mode.

In this work, we adopt the BHF theory for NM,
which is based on realistic many-body forces that describe accurately
nucleon scattering data in free space and the properties of the deuteron.
Moreover, the BHF approach enables the derivation of the properties
of NM at nuclear saturation density
consistent with experiments
\cite{Li08a,Kohno13,Fukukawa15,Wei20,Burgio21}.
For QM, we adopt the Dyson-Schwinger-equation quark model
\cite{Chen11,Chen15},
which can simultaneously address both confinement and
dynamical chiral symmetry breaking \cite{Roberts94,Alkofer01}.
We employ the Gibbs phase transition between the hadron and
deconfined quark phase \cite{Glendenning92,Chen11},
which determines a range of baryon densities
where hadron and quark phase coexist.
In this framework,
the maximum masses of the pure NSs and HSs fulfill the two-solar-mass constraint
\cite{Demorest10,Antoniadis13,Fonseca16,Cromartie19}.

This work is organized as follows.
In Sec.~\ref{s:eos} we briefly describe the formalism for the EOSs, i.e.,
the BHF theory for the hadron phase and the DSEs for the quark phase.
In Sec.~\ref{s:osc} we introduce the TOV and
the eigenvalue equations for the NROs of NSs.
Numerical results are given in Sec.~\ref{s:res},
and we draw the conclusions in Sec.~\ref{s:end}.
We use natural units $c=\hbar=1$ throughout the paper.

%------------------------------------------------------------------------------
\section{Equation of state}
\label{s:eos}

\subsection{Nuclear matter}

The BHF many-body theory \cite{Jeukenne76,Baldo99}
is used to describe the NM in the interior of NSs.
It can reproduce NM properties near the saturation density
with a quite good accuracy \cite{Wei20,Burgio21}.
We only provide here a brief overview of the formalism,
and refer to the various indicated references for full details.
The essential ingredient of this approach is the interaction matrix $G$,
which satisfies the following equations
\be
  G(\rho,\xp;E) = V + V \, \sum_{1,2}
 \frac{\ket{12} (1-n_1)(1-n_2) \bra{12}}
 {E - e_1-e_2 +i0} G(\rho,\xp;E) \:
\label{eq:BG}
\ee
and
\be
 U_1(\rho,\xp) = \sum_2 n_2
 \expval{G(\rho,\xp;e_1+e_2)}{12}_a \:,
\label{eq:uk}
\ee
where $n_i(k)$ is a Fermi distribution,
$\xp \equiv \rho_p/\rho$ is the proton fraction,
and $\rho_p$ and $\rho$ are the proton and the total nucleon number densities,
respectively.
$E$ is the starting energy and
$e_i(k) \equiv k^2\!/2m_i + \Re U_i(k)$ is the single-particle energy.
The multi-indices $1,2$ denote in general momentum, isospin, and spin.
The energy density of NM can then be expressed as
\begin{equation}
 \eps_N =
% \sum_{i=n,p} 2\sum_k n_i(k) \qty( {k^2\over 2m_i} + {1\over 2}U_i(k) )
 \sum_1 n_1(k) \qty( \frac{k^2}{2m_1} + \frac{1}{2}U_1(k) )
\:.
\label{eq:f}
\end{equation}

Thus, the nucleon-nucleon interaction potential $V$ is the only necessary input
in the calculation process.
In this work, we adopt the
Argonne $V_{18}$ (V18) \cite{Wiringa94}
and
Bonn-B (BOB) \cite{Machleidt87,Machleidt89}
potentials,
supplemented with compatible microscopic three-body forces
\cite{Grange89,Zuo02,Li08a,Li08b}.
This is a common prescription adopted in the BHF approach,
and allows to reproduce correctly the saturation point
of symmetric NM and related properties.

In order to obtain the EOS,
we impose cold, neutrino-free, charge neutral, and catalyzed matter
consisting of neutrons, protons, and leptons ($e^-,\mu^-$)
in beta equilibrium due to weak interactions.
Therefore, the energy density of NM can be expressed as
\bal
 \eps(\rho_n,\rho_p,\rho_e,\rho_\mu) =&~
 (\rho_nm_n+\rho_pm_p)  + \eps_N(\rho_n,\rho_p)
\nonumber \\&
 + \eps_e(\rho_e) + \eps_\mu(\rho_\mu)
\:,
\label{eq:ea}
\eal
where $\eps_{e,\mu}$ are the energy densities of electrons and muons,
and $m_{n,p}$ are the masses of neutrons and protons, respectively.

Furthermore, a quadratic dependence on the proton fraction
is well fulfilled
\cite{Bombaci91,Bombaci94,Zuo04,Burgio10,Li21},
\be
 \eps_N(\rho_n,\rho_p) = \eps_\text{SNM}(\rho)
 + (1 - 2\xp)^2 \eps_\text{sym}(\rho) \:,
\label{B/A}
\ee
being $ \eps_\text{sym}(\rho)$ the symmetry energy density,
\be
 \eps_\text{sym}(\rho) = \eps_\text{PNM}(\rho) - \eps_\text{SNM}(\rho) \:.
\label{e:sym}
\ee
Therefore, for the treatment of beta-stable matter,
it is only necessary to calculate the energy densities for
symmetric nuclear matter (SNM) and pure neutron matter (PNM).
For practical use, we employ the convenient empirical parametrizations
given in Refs.~\cite{Lu19,Wei20}.  % Liu22 Li08b
We have shown in \cite{Li21} that going beyond the parabolic approximation
affects the results for NS structure only in a very marginal way.

The various chemical potentials
of the particle species $i=n,~p,~e,~\mu$
can be computed from the energy density, Eq.~(\ref{eq:ea}),
\be
 \mu_i = {\frac{\partial \eps}{\partial \rho_i}} \:,
\ee
and this allows to solve the equations for beta-equilibrium,
\bal
 \mu_p + \mu_e &= \mu_n \equiv \mu_B  \:,\
%\\
 \mu_e = \mu_\mu   \:,
\label{e:beh}
\eal
together with charge neutrality,
\be
% \sum_i \rho_i q_i = 0 \:,
 \rho_p - \rho_e - \rho_\mu \equiv \rho_C = 0 \:,
\label{e:cnh}
\ee
for the equilibrium composition $\rho_i$ at fixed baryon density
$\rho_B=\rho=\rho_p+\rho_n$.
Finally the EOS is given by
\be
 p(\eps) = \rho_B^2 \frac{ d }{ d \rho_B}
 \frac {\eps(\rho_i(\rho_B))} {  \rho_B}
 = \rho_B  \frac{ d \eps }{d \rho_B} - \eps
 = \rho_B \mu_B - \eps \:.
\ee

The BHF approach provides only the EOS for the bulk matter region
$\rho\gtrsim 0.1\fm3$
without cluster formation, and therefore it has to be
joined with a low-density crust EOS.
In this paper we adopt the Shen2020 EOS \cite{Shen20},
which belongs to the class of so-called unified EOSs,
and is frequently used
for the simulations of core-collapse supernova and NS mergers.
%{by Negele-Vautherin \cite{Negele71} in the medium-density regime,
%and the ones by Baym-Pethick-Sutherland \cite{Baym71}
%and Feynman-Metropolis-Teller \cite{Feynman49} for the outer crust.}
We notice that the high-mass domain that we are mainly interested in,
is in any case hardly affected by the structure
of this low-density transition region \cite{Burgio10}.
The choice of the crust model can influence the predictions of
radius and related deformability to a small extent,
of the order of $1\%$ for $R_{1.4}$ \cite{Burgio10,Baldo14b,Fortin16},
which is negligible for our purpose.
Even neglecting the crust completely,
NS radius and deformability do not change dramatically \cite{Tsang19}.

Due to the absence of strict theoretical and observational constraints
on how to join the core and crust EOSs,
we use the simplest way,
a continuous transition
at the point with the same pressure and energy density.
The possible influence of the core-crust transition construction
on the NROs will be discussed in following.

%------------------------------------------------------------------------------
\subsection{Quark matter}

As in Ref.~\cite{Luo19},
we adopt the Dyson-Schwinger model (DSM) \cite{Chen11}
to describe the deconfined quark phase,
which provides a continuous approach to quantum chromodynamics (QCD).
The fundamental quantity of the DSM is the quark propagator $S(p;\mu)$
at finite chemical potential $\mu$,
which satisfies the Dyson-Schwinger equation
\bal\label{e:dse}
 & S(p;\mu)^{-1} = Z_2 \left[ i \bm{\gamma} \cdot \bm{p}
 + i \gamma_4(p_4 + i\mu) + m_q \right] + \Sigma(p;\mu) \:
\eal
with the renormalized self-energy expressed as
\bal\label{dssigma}
 \Sigma(p;\mu) = Z_1 g^2(\mu) \!\! \int \!\!
 \frac{d^4 q}{(2\pi)^4} D_{\rho\sigma}(k;\mu)
 \frac{\la^a}{2} \gamma_\rho S(q;\mu)
%\nonumber\\\times
 \frac{\la^a}{2} \Gamma_\sigma(q,p;\mu) \:,
\eal
where $D_{\rho\sigma}(k\equiv p-q;\mu)$ is the full gluon propagator,
$\Gamma_\sigma(q,p;\mu)$ is the full quark-gluon vertex, and
$Z_1$ and $Z_2$ are the quark-gluon vertex and quark wavefunction
renormalization constants.
Moreover, $\la^a$ are the Gell-Mann matrices,
and $m_q$ is the current-quark bare mass.
Knowing the quark-gluon vertex and gluon propagator,
one can solve the equation and obtain the quark propagator.
In our work,
the so-called rainbow approximation and a
chemical-potential-modified Gaussian-type effective interaction
\cite{Chen11,Luo19} are adopted,
see Ref.~\cite{Chen11} for details.

The EOS for cold QM is obtained following Refs.~\cite{Chen08,Klahn09}.
The quark number density, pressure, and energy density
for each quark flavor at zero temperature can be obtained as
\bal
 \rho_q(\mu_q) &= 6\int\!\! \frac{d^4p}{(2\pi)^4}
 \mathop{\text{tr}_D} \left[-\gamma_4 S_q(p;\mu_q) \right] \:,
\label{e:dsrho}
\\
  p_q(\mu_q) &= p_q(\mu_{q,0}) + \int_{\mu_{q,0}}^{\mu_q} d\mu \rho_q(\mu) \:,
\label{e:dsp}
\\
  \eps_q(\mu_q) &= -p_q(\mu_q) + \mu_q \rho_q(\mu_q) \:.
\label{e:dsed}
\eal
The baryon chemical potential and total baryon number density
in the quark phase are
\bal
\label{e:bcpq}
 & \mu_B = \mu_u + 2\mu_d \:,
\\
\label{e:bndq}
 & \rho_B = \frac{\rho_u+\rho_d+\rho_s}{3} \:,
\eal
and the total pressure and energy density are given
by summing contributions from all quark flavors
and those from electrons and muons.
The pressure of QM at zero density is determined by a
phenomenological bag constant \cite{Wei17},
\be
\label{BDS}
 B_\text{DS} = -\!\!\sum_{q=u,d,s} p_q(\mu_{q,0}) \:,
\ee
which is set to $90\mfm$ \cite{Chen12,Chen15,Wei17}.

The beta equilibrium and charge neutrality in the pure quark phase
are expressed as
\bal
\label{e:beq}
 & \mu_d = \mu_u + \mu_e = \mu_u + \mu_\mu = \mu_s \:,
\\
 & \rho_C = \frac{2\rho_u-\rho_d-\rho_s}{3} - \rho_e - \rho_\mu = 0 \:.
 \label{e:cnq}
\eal
In this work we adopt the Gibbs construction \cite{Glendenning92,Chen11}
for the phase transition between hadron phase and quark phase.
In combination with the respective beta-equilibrium conditions
Eqs.~(\ref{e:beh}) and (\ref{e:beq}),
the chemical and mechanical equilibrium in the mixed phase are expressed as
\bal\label{chemical_equilibrium}
      &\mu_{B,N} = \mu_{B,Q}  \:,
\\
      &\mu_{e,N} = \mu_{e,Q}  \:,
\\
  p_N(&\mu_e,\mu_B) = p_Q(\mu_e,\mu_B) = p_M(\mu_e,\mu_B) \:,
\eal
where the subscripts ``N", ``Q", and ``M"
represent NM, QM, and the mixed phase, respectively.
In the mixed phase,
the local charge neutrality conditions Eqs.~(\ref{e:cnh}) and (\ref{e:cnq})
are replaced by the global condition
\be\label{chi}
 \chi\rho_{C,Q} + (1-\chi)\rho_{C,N} = 0 \:,
\ee
where $\chi$ is the volume fraction of QM in the mixed phase.
Consequently,
the baryon number density $\rho_{B,M}$ and energy density $\eps_M$
of the mixed phase can be determined as
\bal\label{energy_density}
 \rho_{B,M} &= \chi\rho_{B,Q} + (1-\chi)\rho_{B,N} \:,
 \\
 \eps_M &= \chi\eps_Q + (1-\chi)\eps_N \:.
\eal
Specifically, in the Gibbs construction
the pressure and energy density in HSs
are continuous functions of the baryon density,
at variance with the Maxwell phase transition \cite{Pereira18}.

%------------------------------------------------------------------------------
\section{Neutron Stars}
\label{s:osc}

\subsection{Hydrostatic equilibrium structure}

\iffalse
Due to the strong gravitational field in NSs,
their structure and dynamical evolution are ruled
by the Einstein equation in general relativity,
\be\label{field}
 R_{\mu\nu} - \frac{1}{2}g_{\mu\nu}R = 8\pi G T_{\mu\nu} \:,
\ee
where $R_{\mu\nu}$ is the Ricci tensor,
$R$ is the Ricci scalar, and
$G$ is the gravitational constant.
The energy-momentum tensor is
\be\label{tensor}
 T_{\mu\nu} = pg_{\mu\nu} + (p+\eps)u_\mu u_\nu \:,
\ee
where $g_{\mu\nu}$ is the metric tensor,
$p$ is the pressure,
$\eps$ is the energy density,
and $u_\mu$ is the four-velocity.
For simplicity, we consider static spherically symmetric stars,
described by the Schwarzschild metric
\cite{Chandrasekhar64}
\fi

The general static spherically-symmetric metric
which describes the geometry of a static NS can be written as
\be\label{e:ds2}
 ds^2 = e^{\nu(r)}dt^2 - e^{\la(r)}dr^2
 - r^2(d\theta^2+\sin^2\!\theta d\phi^2) \:,
\ee
where $e^{\nu(r)}$ and $e^{\la(r})$ are metric functions.
The TOV equations \cite{Oppenheimer39,Tolman39}
obtained from the Einstein field equation for the metric are
\bal\label{dpdr}
 \frac{dp}{dr} &= G\frac{(\eps+p)(m+4\pi r^3p)}{r^2(2Gm/{r}-1)} \:,
\\
 \frac{dm}{dr} &= 4\pi r^2\eps \:,
\eal
and the corresponding metric functions
\bal\label{A}
 e^{\la(r)} &= (1-2Gm/r)^{-1} \:,
\\
 \nu(r) &=
 -2G \int_r^\infty\!\! dr' \frac{e^{\la(r')}}{r'^2}
 \left( m + 4\pi r'^3 p \right) \:.
\eal
The boundary conditions
$m(r=0)=0$, $p(r=0)=p_c$, and $p(R)=0$,
where $p_c$ is the central pressure,
lead to equilibrium configurations in combination with the EOS of the NS matter,
thus obtaining radius $R$ and mass $M=m(R)$
of a NS for a given central pressure or density.

%------------------------------------------------------------------------------
\subsection{Non-radial oscillations}

Thorne developed a complete theory for NROs of NSs
from the Einstein field equations \cite{Thorne67}.
The perturbation of the fluid in the star is described
by the Lagrangian displacement vector $\xi^\al$ in terms of
the dimensionless perturbation functions $W(r)$ and $V(r)$,
\bal\label{e:xi}
 \xi^r      &= r e^{-\la/2} W Y^l_m \,e^{i\om t} \:,
\nonumber\\
 \xi^\theta &= -V \partial_\theta Y^l_m \,e^{i\om t} \:,
\nonumber\\
 \xi^\phi   &= -(\sin\theta)^{-2} V \partial_\phi Y^l_m \,e^{i\om t} \:,
\eal
where $Y^l_m(\theta,\phi)$ are the usual spherical harmonics,
and the eigenvalue $\om = 2\pi f$ is the frequency of the NRO.
The eigenfunctions $W(r)$ and $V(r)$ are determined by the pulsation equations.
The full NRO equations can be found in the literature
\cite{Thorne67,Detweiler85}.

As we mainly focus on the $g$-mode oscillations,
in this work we consider the relativistic Cowling approximation,
which disregards the perturbations in the metric \cite{Cowling41}.
The validity of the Cowling approximation has been confirmed in
Refs.~\cite{Yoshida97,Chirenti15,Xu17,Zhao22},
which find relatively small deviations of the $g$-mode oscillation frequencies
compared to the full solutions.
As for the other two modes,
this approximation yields about 10--30\% accuracy
of the $f$-mode frequencies
and about $20\%$ for the $p$ mode.

The oscillation equations in Cowling approximation are
\bal\label{e:after1}
 r \frac{dW}{dr} &= \left( \frac{g}{c_s^2} - 3 \right) W
 + e^{\la/2} \left( \frac{\om^2 r^2}{c_s^2 e^\nu} - l(l+1) \right) V
\:,\\
 r \frac{dV}{dr} &= e^{\la/2}\left( \frac{N^2}{\om^2} - 1 \right) W
 + \left( 2g + \frac{N^2 r^2}{g e^{\nu-\la}} - 2 \right) V \:,
\label{e:after2}
\eal
with $g \equiv -r(dp/dr) / (p+\eps)$,
and $N$ being the Brunt-V\"ais\"al\"a (BV) frequency defined as
\be\label{e:bvf}
 N^2 \equiv \frac{g^2}{r^2}
 \left( \frac{1}{c_e^2}-\frac{1}{c_s^2} \right) e^{\nu-\la} \:.
\ee
In the Newtonian approximation $N$ is the frequency of the perturbed
fluid elements forced by buoyancy to perform harmonic oscillations.

The BV frequency depends on the difference of the inverse squared
equilibrium speed of sound $c_e$
and adiabatic speed of sound $c_s$,
caused by the deviation from beta-equilibrium during fast enough oscillations.
The former is defined as the derivative of the EOS in beta equilibrium,
\be
 c_e^2 \equiv \frac{dp/dr}{d\eps/dr} = \frac{dp}{d\eps} \:,
\ee
while,
assuming that all weak reactions are slow compared to the oscillation timescale,
the squared adiabatic speed of sound is
\be\label{e:cs}
 c_s^2 \equiv \Big( \frac{\partial p}{\partial \eps} \Big)_{S,Y_i}
 = \frac{\Delta p}{\Delta \eps}
\:,
\ee
where $S$ and $Y_i$ denote entropy and the particles fractions
affected by weak reactions, respectively.
It is related to the adiabatic index of the fluid,
\be\label{e:gamma}
 \gamma = \left( 1 + \frac{\eps}{p} \right) c_s^2 \:,
\ee
which drives the response of the stellar material to pulsational perturbations.

However, one should note that Eq.~(\ref{e:cs}) is only adequate
in the pure hadron/quark phases.
In the mixed phase,
one also needs to consider the transition between NM and QM,
as well as the independent expansion/contraction of NM/QM
during the oscillation.
The related conversion rate/time scale between hadron and quark phases
is still an open question,
and current calculations are model dependent.
Following the argument of \cite{Lugones21},
the preferred conversion timescale would be slow at the hadron-quark interface,
although a rapid timescale cannot be discarded \cite{Rodriguez21,Ranea22}.

Therefore, herein we neglect the phase transition
between NM and QM in the mixed phase,
keeping all particles fractions $Y_i$ constant
in NM and QM separately.
Note that this means that the strong chemical equilibrium
between NM and QM is violated, i.e.,
the isospin chemical potentials in NM and QM are different during oscillations.
The volume fraction of QM in the mixed phase changes during the oscillation,
to keep the pressure equilibrium between NM and QM.
Accordingly, the squared adiabatic speed of sound in the mixed phase
is modified as \cite{Jaikumar21}
\be\label{e:cs2}
 c_s^{-2} = \frac{\chi'\eps'_Q + (1-\chi')\eps'_N - \eps}{p'-p} \:,
\ee
where $p$ and $\eps$ are the pressure and energy density in beta-equilibrium,
$p'=p+\Delta p$ is the pressure during oscillations,
$\eps'_Q$, $\eps'_N$ are the energy densities of QM and NM
during oscillations,
and $\chi'$ is determined by $p'_Q=p'_N =p'$ during oscillations.

In this work, we focus on the influence of the core on the NROs,
as in recent literature \cite{Jaikumar21,Constantinou21,Zhao22}.
For simplification,
we will make the approximation that $c_s=c_e$ in the crust,
and correspondingly the BV frequency $N_\text{crust}=0$.
The influence of the crustal $N$ on the core modes of NROs is small,
which will be discussed in the following section.

In order to determine the eigenfrequencies of oscillation,
one also needs boundary conditions \cite{McDermott83}, which,
in the NS center, are given by
\be\label{e:r0}
 W(0) + lV(0) = 0 \:,
\ee
while at the outer surface of the star,
the Lagrangian perturbation of the pressure should vanish,
\be\label{e:rR}
 \Delta p(R) = \gamma p
 \left[ -e^{-\la/2} \Big(r\frac{dW}{dr} + 3W\Big) - l(l+1) V \right](R) = 0 \:.
\ee
Due to the homogeneousness of the oscillation equations
Eqs.~(\ref{e:after1},\ref{e:after2}),
one can impose arbitrarily
$W(r=0)=1$ and $V(r=0)=-1/2$
at the center.
Then integrating Eqs.~(\ref{e:after1},\ref{e:after2})
from center to the surface with the boundary conditions
yields the discrete eigenfrequencies $\om_i$
and the eigenfunctions $W_i,V_i$.
For the quadrupole $(l=2)$ oscillations of pure NSs and HSs,
they can be ordered as
$\om_{g_n}<\ldots<\om_{g_1}<\om_f<\om_{p_1}<\ldots<\om_{p_n}$,
where $n$ is the number of nodes.

%-------------------------------------------------------------------------------
\subsection{GW damping time}

The damping time of oscillations through GW emission is
\be\label{e:tau}
 \taugw = \frac{2E}{P_\text{GW}} \:,
\ee
where $E$ is the total energy stored in the oscillation,
and $P_\text{GW}$ is the power of the GW radiation released by the mode.

The energy per radial distance
of an eigenmode in Cowling approximation
is given by \cite{McDermott83}
\be\label{e:osenergy}
 \frac{dE}{dr} = \frac{\om^2}{2} (p+\eps) e^{(\la-\nu)/2}
 r^4 \left[ W^2 + l(l+1)V^2 \right] \:,
\ee
and its power can be estimated as \cite{Reisenegger92}
\be\label{e:pgw}
 P_\text{GW} = \frac{G(l+1)(l+2)}{8\pi (l-1)l}
 \left[ \frac{4\pi \om^{l+1}}{(2l+1)!!}
 \int_0^R dr\, r^{l+2} \deps \right]^{2} \:,
\ee
where $\deps$ is the Eulerian perturbation of the energy density,
in Cowling approximation given as \cite{McDermott83}
\bal
 \deps = & -(p+\eps)
 \left[ e^{-\la/2}\left(3W + r \frac{dW}{dr} \right) + l(l+1) V \right]
\nonumber\\&
 - r \frac{d\eps}{dr} e^{-\la/2} W \:.
\label{e:deps}
\eal
It has been pointed out \cite{Reisenegger92}
that the different terms in this equation are of different signs
and tend to cancel each other,
which renders a reliable numerical evaluation very delicate.
We will come back to this problem later.

The radiation power is related to the GW strain (metric perturbation),
which in quadrupole approximation and transverse-traceless (TT) gauge
is \cite{Thorne80,Szczepanczyk21}
\be\label{e:h}
 h_{ij}^\text{TT}(t,D) =
 \frac{2G}{D} \ddot{Q}_{ij}^\text{TT}(t-D/c) \:,
\ee
where $i,j=1,2,3$ are the indices in Cartesian coordinates,
$D$ is the distance to the source,
$G$ is the gravitational constant,
the two dots represent the second time derivative,
and the traceless quadrupole moment is
\be\label{Q}
 Q_{ij}^\text{TT}(t) = \int d^3\xv\, \eps(t,\xv)
 \Big( x_i x_j - \frac13 \delta_{ij}|\xv|^2 \Big) \:.
\ee
The metric perturbation tensor can be decomposed as
\be\label{h1}
 \mathbf{h}^\text{TT} = h_+ \et_+ + h_\times \et_\times \:,
\ee
where $\et_+$ and $\et_\times$
are the unit tensors of plus and cross polarization.

In the case of a symmetric metric,
the cross polarization $h_\times$ is zero,
$Q_{ij}$ has only diagonal components
$Q_{11} = Q_{22} = -\frac12 Q_{33}$,
and the GW strain can be calculated by \cite{Finn90}
\be
 h_+ = \frac{3G\sin^2\al}{2D} \ddot{Q}_{33} \:,
\ee
where $\al$ is the inclination angle.
In this work, we choose $\sin\al=1$.
After some derivation, $\ddot{Q}_{33}$ can be written as
\bal
 \ddot{Q}_{33}
 &= \left| \ddot{Q}_{33}\right| e^{-i\om t}
\nonumber\\
 &= \int dr\, r^2 \sin\theta d\theta d\phi\,
 r^2 \left(\sin^2\theta - \frac13 \right)
 \frac{d^2 \deps}{dt^2}
\:,
\eal
and thus
\be
 \left|\ddot{Q}_{33}\right| =
 \frac{4\sqrt{\pi/5}}{3} \om^2 \int dr\, r^4 \deps(r) \:,
\ee
where $\deps$ is given by Eq.~(\ref{e:deps}).
So the amplitude of the GW strain can be rewritten as
\be
 \left|h_+\right| = \frac{3G\left|\ddot{Q}_{33}\right|}{2D} \:,
\label{e:hp}
\ee
and depends, as the radiation power, on $\deps(r)$.

%------------------------------------------------------------------------------
\section{Numerical results}
\label{s:res}

\begin{figure}[t]%.............................................................
\centerline{\includegraphics[scale=0.75]{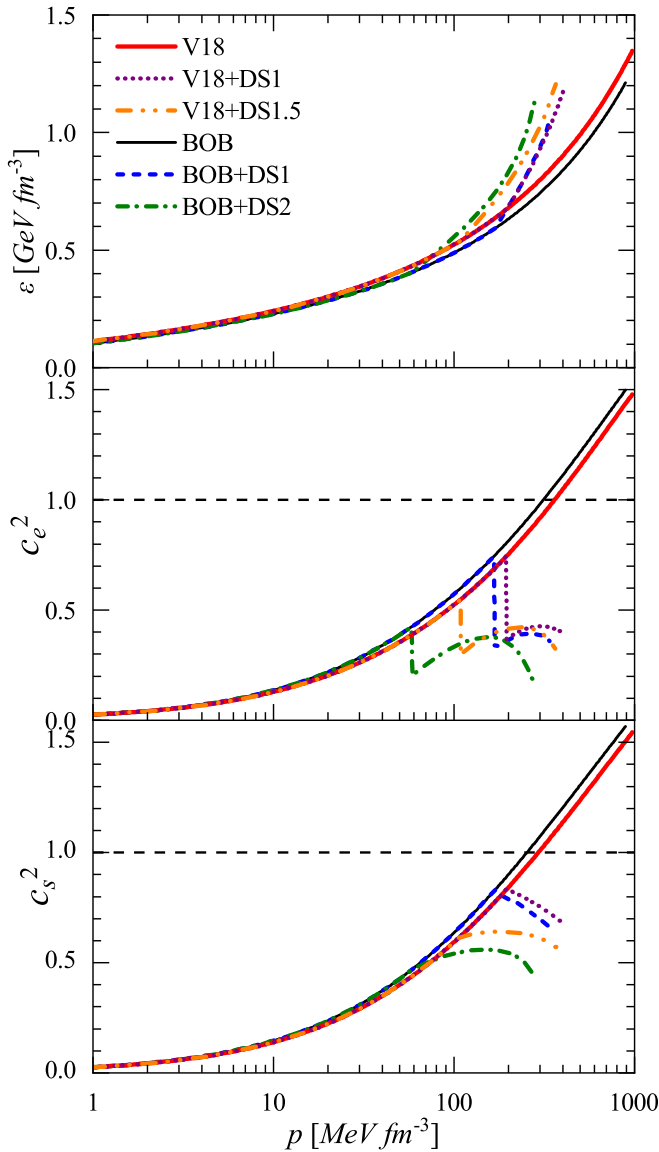}}
\vskip-3mm
\caption{
The energy density (upper panel),
squared equilibrium speed of sound (central panel),
and squared adiabatic speed of sound (lower panel) of NS matter
as functions of pressure with different EOSs.
All curves end at the $\mmax$ configuration for the proper EOS.
See the text for a detailed description of the notation.
}
\label{f:eos}
\end{figure}%..................................................................

\begin{figure}[t]%.............................................................
\centerline{\includegraphics[width=0.5\textwidth]{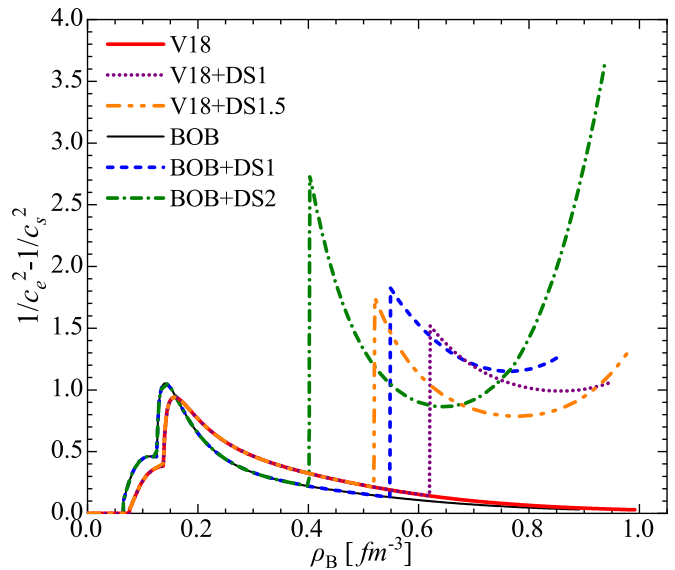}}
\vskip-3mm
\caption{
Difference of squared inverse sound speeds
$1/c_e^2-1/c_s^2$ vs. baryon number density.
All curves end at the $\mmax$ configuration for the proper EOS.
}
\label{f:delta_c}
\end{figure}%..................................................................

%------------------------------------------------------------------------------
\subsection{EOS and equilibrium structure of neutron stars}

As discussed above, we use the BHF EOS with BOB or V18 potential
for the pure NM.
Regarding the QM EOS in HSs,
there is a free model parameter $\al$,
which represents the strength of the
in-medium modification of the Gaussian-type effective interaction in the DSM.
Here we choose $\al = 1,2$ and $\al = 1,1.5$
in combination with the BHF BOB and V18 EOSs, respectively,
labeled as BOB/V18+DS$\al$,
to satisfy both the requirements of $\mmax>2\ms$ and causality,
as discussed in detail in \cite{Chen11,Luo19}.

The energy density, squared equilibrium speed of sound,
and squared adiabatic speed of sound as functions of the pressure
are shown in Fig.~\ref{f:eos}.
The colors/types of curves refer to the different combinations
of BHF and DSM EOS.
The EOS of the mixed phase (broken lines) is generally softer
than that of the pure NM (solid lines),
whereas pure QM emerges at too high densities
that cannot be reached in HSs in our approach.
However, the QM onset density is strongly dependent on the theoretical model
adopted for the description of QM.

We notice that both sound speeds in pure NM
become superluminal at high densities
(but quite close to the $\mmax$ configuration),
due to the non-relativistic character of the BHF theory.
However, in our model,
the phase transition to QM occurs always at lower density,
and consequently causality is never violated.
The equilibrium speed of sound $c_e$ is particularly sensitive
to the composition of the matter.
One can find a small discontinuity at $p\approx5\mfm$
due to the onset of muons
(hardly visible because of the scale),
and a sharp discontinuous drop at the phase transition point,
due to the appearance of QM.
Unlike $c_e$, the adiabatic speed of sound $c_s$ is always continuous
with increasing pressure,
independently of the emergence of muons or quarks.
This is because it is defined with fixed particle composition, Eq.~(\ref{e:cs}).
Similar results were obtained in Ref.~\cite{Zhao22},
using different NM and QM EOSs.

In Fig.~\ref{f:delta_c} we show the difference of squared inverse sound speeds
$1/c_e^2-1/c_s^2$ as function of the baryon number density in NS matter.
This difference determines the profile of the local oscillation frequency,
i.e., the BV frequency, Eq.~(\ref{e:bvf}).
As discussed above, we approximate $c_s=c_e$ and $N=0$ in the crust,
therefore the difference is zero below the core-crust transition density.
In the core with the pure BHF EOS,
$c_s$ and $c_e$ increase with density,
and therefore $1/c_e^2-1/c_s^2$ decreases.
There is a first spike around $\rho_B \simeq 0.15\fm3$
due to the muon onset,
and at larger densities a second much sharper spike due to the appearance of QM.
In the mixed phase the difference decreases and then increases again
when approaching the pure quark phase.
We stress that the (sharp) change of $1/c_e^2-1/c_s^2$
is mainly due to the (sharp) change of $c_e$,
which is very sensitive to the change of particle species,
see Fig.~\ref{f:eos}.

Furthermore we notice that,
whereas the differences in the purely hadronic phase become very small
and rapidly approach zero,
those in the mixed phase are larger and can rise very quickly.
This behavior already hints to larger values of the $g$-mode frequencies
and a strong dependence on the QM EOS
in hybrid stars compared to the purely hadronic stellar configurations.

\begin{figure}[t]%.............................................................
\centerline{\includegraphics[width=0.5\textwidth]{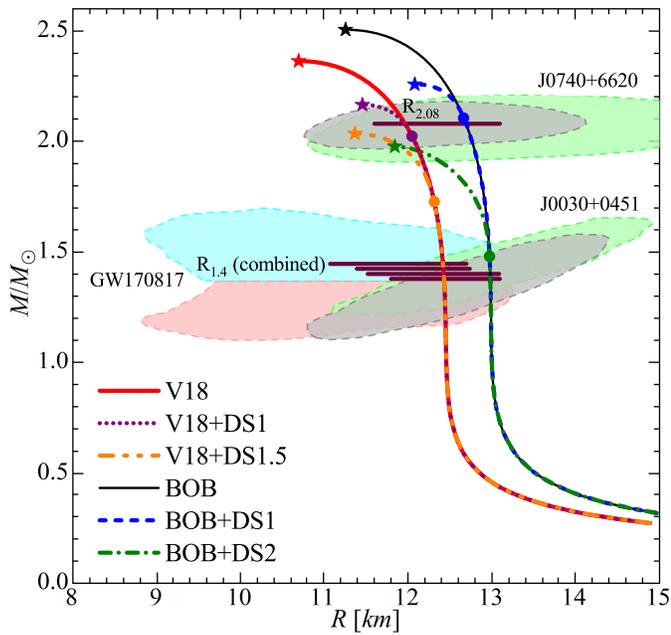}}
\vskip-2mm
\caption{
The mass-radius relations of NSs obtained with different EOSs.
Full dots indicate the bifurcation points of pure NSs and HSs.
The horizontal black bars indicate the limits on
$R_{2.08}$ and $R_{1.4}$
obtained in the combined NICER+GW170817 data analyses
of \cite{Miller21,Pang21,Raaijmakers21}.
The maximum-mass configurations are indicated by star symbols.
}
\label{f:mr}
\end{figure}%..................................................................

\begin{figure}[t]%.............................................................
\vskip-1mm
\centerline{\includegraphics[width=0.51\textwidth]{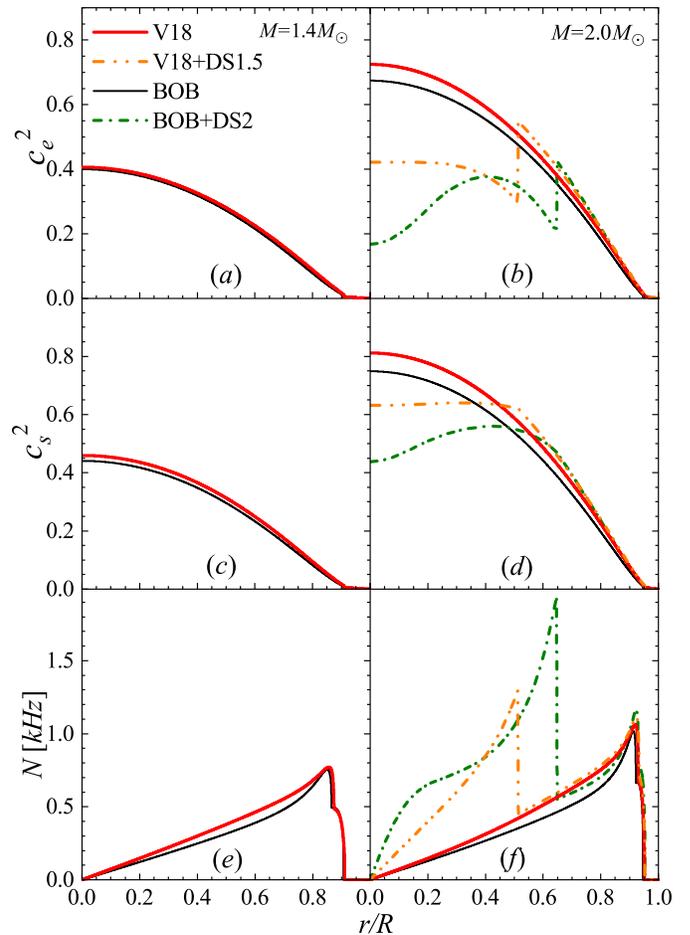}}
\vskip-2mm
\caption{
Squared equilibrium speed of sound (upper panels),
squared adiabatic speed of sound (central panels),
and BV frequency $N$ (lower panels)
in NSs with $1.4\ms$ (left panels)
and $2.0\ms$ (right panels),
for various EOSs.
%\hj{bigger font}
}
\label{f:internal}
\end{figure}%..................................................................

\begin{figure}[t]%.............................................................
\centerline{\includegraphics[scale=0.42]{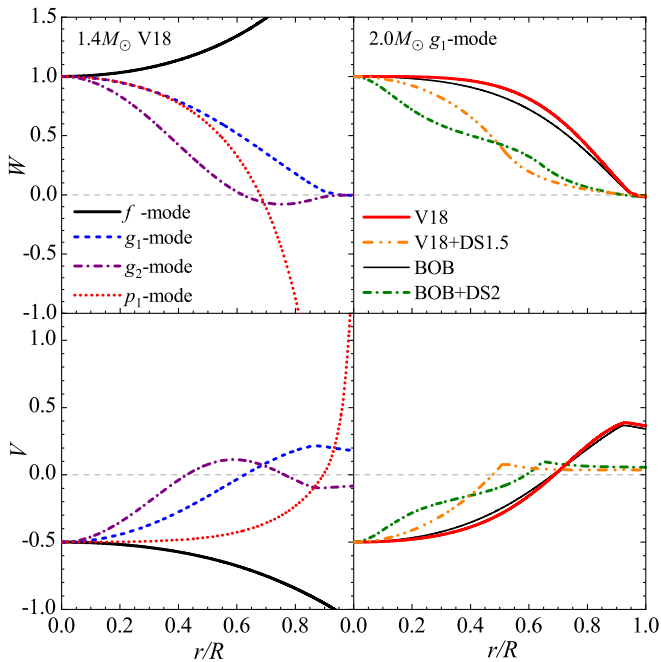}}
\vskip-2mm
\caption{
The radial displacement perturbation $W$ (upper panels)
and the tangential displacement perturbation $V$ (lower panels)
of the first four eigenmodes of NROs
for a $1.4\ms$ NS with the $V18$ EOS (left panels)
and of the $g_1$ mode for a $2.0\ms$ NS with various EOSs (right panels).
%\hj{bigger font}
}
\label{f:internal2}
\end{figure}%..................................................................

%\begin{figure}[t]%.............................................................
%\vskip-0mm
%\centerline{\includegraphics[width=0.51\textwidth]{osenergy}}
%\vskip-4mm
%\caption{The radial displacement perturbation $W$ (upper panels),
%the tangential displacement perturbation $V$ (central panels),
%and the radial gradient of oscillation energy (lower panels)
%of the $g_1$ mode with various %EOSs for a $2\ms$ star.}
%\label{f:osenergy}
%\end{figure}%..................................................................

\begin{figure}[t]%.............................................................
\centerline{\includegraphics[width=0.5\textwidth]{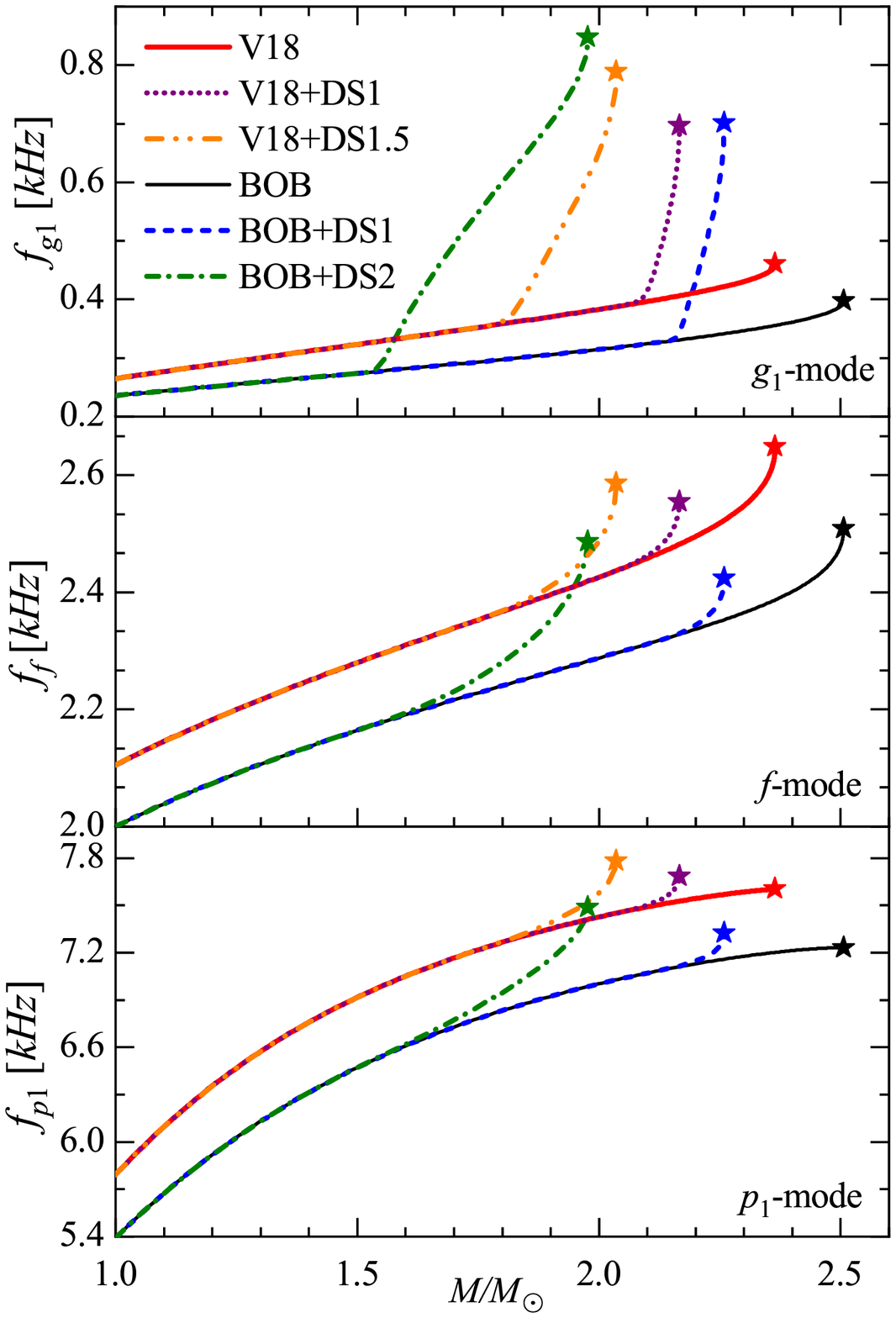}}
\vskip-4mm
\caption{
$g_1$-, $f$-, and $p_1$-mode frequencies vs mass $M$
for various EOSs.
%\hj{[],blue curves}
}
\label{f:f-m}
\end{figure}%..................................................................

The mass-radius relations of pure NSs and HSs with the various considered EOSs
are shown in Fig.~\ref{f:mr}.
The broken curve segments indicate the hybrid star branches.
One notes that the differences between HSs and pure NSs are even smaller
than the differences between pure NSs with different BHF EOSs.
Therefore, it is difficult to distinguish HSs from pure NSs
with only $M(R)$ observations.
Within our model, pure QM matter does not appear in the core of HSs.

All plotted EOSs
(in particular the V18 models)
fulfill the constraints from present observations on NS mass and radius,
in particular the recent mass-radius results of the NICER mission
for the pulsars J0030+0451 \cite{Riley19,Miller19}
and J0740+6620 \cite{Riley21,Miller21,Pang21,Raaijmakers21}.
The combined (strongly model-dependent) analysis
of both pulsars together with the GW170817 event observations
\cite{Abbott17a,Abbott18}
yields improved limits on the radius
$R_{2.08}=12.35\pm0.75\km$ \cite{Miller21},
but in particular on $R_{1.4}$, namely
$12.45\pm0.65\km$ \cite{Miller21}, %11.8-13.4 (90%) and 12.2-13.1 (68%)
$11.94^{+0.76}_{-0.87}\km$ \cite{Pang21}, and
$12.33^{+0.76}_{-0.81}\km$ or
$12.18^{+0.56}_{-0.79}\km$ \cite{Raaijmakers21},
which are shown as horizontal black bars in the figure.
%for the V18/BOB EOS $R_{1.4}=12.3/12.8\km$
The BHF V18 EOS is well compatible with these constraints
\cite{Wei20,Burgio21,Sun21},
and also its maximum mass $\mmax \approx 2.36\ms$
exceeds the currently known lower limits.
The BOB EOS is stiffer and allows a maximum mass of even $2.51\ms$.
Some theoretical analyses of the GW170817 event
indicate also an upper limit on the maximum mass
of $\sim$2.2--2.4$\,\ms$
\cite{Shibata17,Margalit17,Ruiz18,Rezzolla18,Shibata19},
with which the V18 EOS would be compatible as well.
However, those are very model dependent,
in particular the still to-be-determined temperature dependence of the EOS
\cite{Khadkikar21,Bauswein21,Figura21,Liu22}.

In Fig.~\ref{f:internal} we show the profiles of both the
equilibrium and adiabatic squared speed of sound,
and the BV frequency
in NSs with masses $M=1.4,\,2.0\ms$
for different EOSs.
The speed of sound generally increases from the crust to the center of the NS,
apart from the density region close to the pure quark phase or a sudden drop of $c_e$ due to the onset of QM
(e.g., upper right panel),
which does not take place in the $1.4\ms$ configurations
shown in the left panels.
By definition, the BV frequency vanishes in the center
and increases towards the crust,
exhibiting sharp decreases at the phase transition point or when muons disappear.
We see that in the inner core of HSs,
the BV frequency is much larger than in pure NSs.
The BV frequency is a local property in NSs which cannot be observed directly,
but it is closely related to the global $g$-mode frequency.
Accordingly, one can consider the BV frequency and the $g$-mode oscillations
as useful probes of the appearance of QM.
We will discuss this in detail now.

%------------------------------------------------------------------------------
\subsection{Non-radial oscillations of neutron stars}

In this work, we investigate the quadrupole oscillations $(l=2)$
of both pure NSs and HSs.
For illustration,
we show some typical solutions
for the radial and tangential displacement perturbations
$W(r)$ and $V(r)$ in Fig.~\ref{f:internal2} .
The left panels contain the results of four representative
$f$, $g_1$, $g_2$, and $p_1$ eigenmodes
of a $1.4\ms$ NS with the V18 EOS,
exhibiting the expected number of nodes,
whereas the right panel displays
the $g_1$ mode of $2.0\ms$ NSs with various EOSs.
In this case, for purely hadronic NSs (solid lines),
one can see quite similar oscillation amplitudes nearly independent
of the nuclear EOS,
whereas the amplitudes in HSs (broken lines) decrease more quickly
in the inner core and remain smaller in the outer layers.
Therefore, the oscillation in HSs occurs mainly in the inner core,
and might reveal information about the QM/mixed phase.
The $W(r)$ eigenfunction exhibits a smooth behavior over the entire star's profile,
even at the radial distance corresponding to the mixed phase onset;
this is at variance with $V(r)$,
which shows sudden kinks when QM appears.

Now we turn to the discussion of the NRO frequencies in NSs and HSs,
for $g$, $f$, and $p$ modes.
As known, they are classified according to the different restoring forces
acting in the fluid,
and are characterized by different frequencies.
In Fig.~\ref{f:f-m},
we show the NRO frequencies of NSs as functions of NS mass.
One can see that $f_{g_1}$ of pure NSs lies in the range 0.2--0.4\,kHz,
and increases slowly with the NS mass.
The difference of the eigenfrequencies with BOB and V18 EOSs
is always about 0.1\,kHz,
which is related to the corresponding difference of the BV frequencies,
Fig.~\ref{f:internal},
and in particular to the different NS radii,
Fig.~\ref{f:mr}:
the smaller V18 star is oscillating faster.
The same qualitative correlations are observed for the HSs,
which exhibit much higher eigenfrequencies up to 0.9\,kHz.
Therefore $f_{g_1}$ might be a good observable to distinguish HSs from pure NSs.
Again, our results are similar to those discussed in
Refs.~\cite{Zhao22,Jaikumar21,Kumar23}.

Equivalent features are observed
for the much higher eigenfrequencies of $f$ mode and $p_1$ mode,
above 2\,kHz and 7\,kHz, respectively.
But in this case,
the differences between HSs and pure NSs are not larger
than the difference between pure NSs with different models.
Therefore, they can not be used as good observables
to distinguish HSs from purely hadronic NSs.

The eigenfrequencies of several representative modes
of NSs with $M=1.4,2.0\ms$ and various EOSs
are also listed in Table~\ref{t:f}.
One can see that the frequencies of higher-order $g$ modes decrease
and those of higher-order $p$ modes increase.
These results are qualitatively similar to those recently published in
Ref.~\cite{Jaikumar21,Zhan21}.

\begin{table}[t]%..............................................................
\caption{
The NRO frequencies $f$ (in units of kHz)
of six representative eigenmodes
of NSs with $M =1.4,2.0\ms$ and various EOSs.}
\label{t:f}
\def\myc#1{\multicolumn{1}{l}{#1}}
\def\myt#1{\multicolumn{1}{c}{$\tilde{#1}$}}
\renewcommand{\arraystretch}{1.2}
\begin{ruledtabular}
\begin{tabular}{c|cc|rrrr}
%\vspace{0.1cm}
      &\multicolumn{2}{c|}{$1.4\ms$} & \multicolumn{4}{c}{$2.0\ms$} \\
 Mode & \myc{V18} & BOB & V18 & \myc{V18+DS1.5} & BOB & \myc{BOB+DS2} \\
\hline
 $g_3$ &  0.15 &  0.13 &  0.19 &  0.33 &  0.16 &  0.44 \\
 $g_2$ &  0.20 &  0.18 &  0.25 &  0.39 &  0.21 &  0.57 \\
 $g_1$ &  0.31 &  0.27 &  0.38 &  0.65 &  0.31 &  0.85 \\
 $f$   &  2.25 &  2.14 &  2.43 &  2.49 &  2.29 &  2.49 \\
 $p_1$ &  6.77 &  6.33 &  7.43 &  7.58 &  7.00 &  7.49 \\
 $p_2$ &  8.74 &  8.21 & 10.59 & 10.78 &  9.88 & 10.48 \\
\end{tabular}
\end{ruledtabular}
\end{table}%...................................................................

%------------------------------------------------------------------------------
\subsection{GW emission and damping of $\bf g_1$ mode}

\begin{figure}[t]%.............................................................
\centerline{\includegraphics[width=0.51\textwidth]{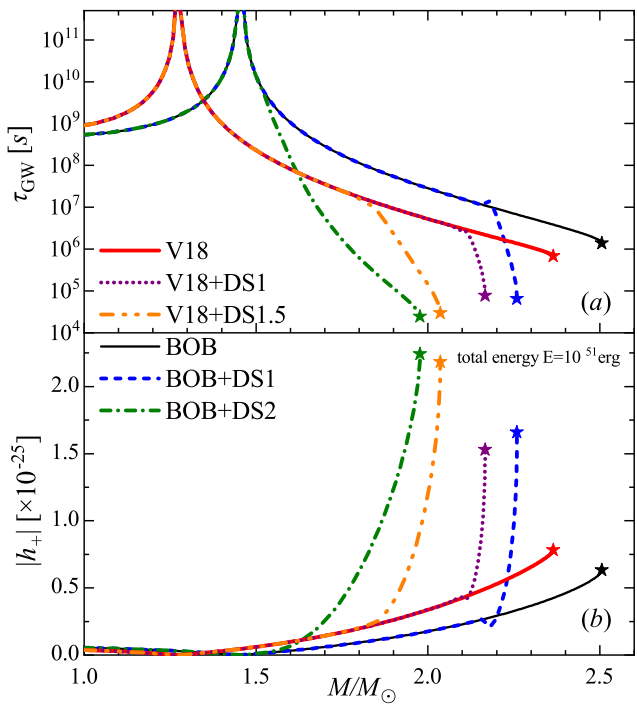}}
\vskip-3mm
\caption{
Properties of $g_1$-mode oscillations vs NS mass $M$ for various EOSs:
(a) The damping time and
(b) the amplitude of the GW strain $|h_+|$
for a total GW energy $E = 10^{51}\,$erg and distance $D = 15\,$Mpc.
}
\label{f:tau-m}
\end{figure}%..................................................................

\begin{figure}[t]%.............................................................
\centerline{\includegraphics[scale=0.6]{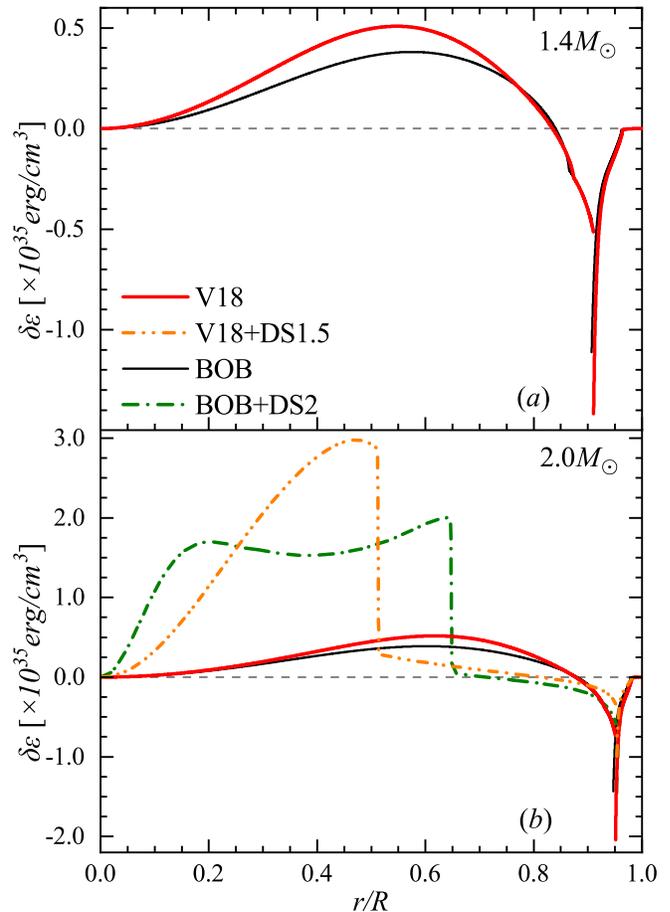}}
\vskip-3mm
\caption{
Eulerian perturbation of the energy density,
Eq.~(\ref{e:deps}),
in NSs with $1.4\ms$ (a)
and $2.0\ms$ (b)
for various EOSs.
The normalization is as in Fig.~\ref{f:tau-m}.
}
\label{f:delta_eps}
\end{figure}%..................................................................

\begin{table*}[t]%.............................................................
\caption{
The minimum detectable energy $E_\text{GW}$, Eq.~(\ref{e:egw}),
(in units of erg)
of $M =1.4,2.0\ms$ NSs
at two representative distances
for various EOSs.
}
\renewcommand{\arraystretch}{1.2}
\begin{ruledtabular}
\begin{tabular}{c|l|cc|cccc}
  &     &\multicolumn{2}{c|}{$1.4\ms$} & \multicolumn{4}{c}{$2.0\ms$}  \\
\hline
Detector   &Distance & V18                & BOB                & V18                & V18+DS1.5          &   BOB              &  BOB+DS2           \\
\hline
LIGO/Virgo & 10\,kpc & $1.5\times10^{46}$ & $1.1\times10^{46}$ & $2.3\times10^{46}$ & $6.8\times10^{46}$ & $1.6\times10^{46}$ & $1.1\times10^{47}$ \\
LIGO/Virgo & 15\,Mpc & $3.5\times10^{52}$ & $2.5\times10^{52}$ & $5.2\times10^{52}$ & $1.5\times10^{53}$ & $3.5\times10^{52}$ & $2.6\times10^{53}$ \\
Einstein   & 10\,kpc & $6.2\times10^{44}$ & $4.5\times10^{44}$ & $9.3\times10^{44}$ & $2.7\times10^{45}$ & $6.3\times10^{44}$ & $4.6\times10^{45}$ \\
Einstein   & 15\,Mpc & $1.4\times10^{51}$ & $1.0\times10^{51}$ & $2.1\times10^{51}$ & $6.1\times10^{51}$ & $1.4\times10^{51}$ & $1.0\times10^{52}$ \\
\end{tabular}
\end{ruledtabular}
\label{t:E_GW}
\end{table*}%..................................................................

In the previous section,
we have pointed out the significant difference of the $g$-mode frequencies
between massive NSs and HSs.
In the following, we investigate a closely related observable,
the damping time of oscillations through GW emission,
Eq.~(\ref{e:tau}).
Before illustrating the results,
we remind that three possible damping mechanisms can act on $g$-mode oscillations, i.e., relaxation toward chemical equilibrium,
viscous damping,
and damping due to emission of GWs.
In Ref.~\cite{Reisenegger92} damping times of the core $g$ modes
(as well as of crustal discontinuity $g$ modes)
of order $10^8$--$10^{11}\,$s were obtained
(but are subject to delicate numerical cancellations).
Thus damping due to emission of GWs is very inefficient,
and the other mechanisms are dominant.

Our results are shown in Fig.~\ref{f:tau-m}(a)
for $g_1$-mode oscillations with various EOSs
as a function of NS mass.
In the low-mass region ($\lesssim1.4\ms$),
we obtain also very large values of the damping time,
and even a divergence at certain masses.
This large value of $\taugw$ is due to the smallness of the integral
for the GW power, Eq.~(\ref{e:pgw}),
caused by cancellations between the individual terms of Eq.~(\ref{e:deps}).
The integral represents a superposition of GWs emitted from perturbations
of the energy quadruple moment at different layers of the NS.
%Compared with the GW wavelength,
%the difference of the distance between the source and the wave,
%i.e., the scale of NSs is negligible.

This is illustrated in Fig.~(\ref{f:delta_eps}),
which shows the Eulerian perturbation
of the energy density, $\deps(r)$, Eq.~(\ref{e:deps}),
in $1.4\ms$ and $2.0\ms$ NSs.
Similar as the perturbation functions $W(r)$ and $V(r)$,
the perturbation of the energy (quadruple moment)
also has nodes at certain radii.
The sign flip of $\deps$ represents a half-period difference
of the oscillation phase of the energy quadruple moment.
Therefore the GWs emitted from domains with different signs of $\deps$
will interfere with each other,
which results in cancellations of the power output.
In our models, we obtain contributions
%to the integral in Eq.~(\ref{e:pgw})
of positive $\deps$ in the inner core
and negative $\deps$ in the outer layers
of the same order of magnitude.
As a result, we obtain large values of the damping time $\taugw$,
and even a divergence at a certain mass.
We stress that the (small) values of the GW power are quite model dependent,
which could be greatly influenced by various approximations,
an so do the values of the damping time $\taugw$ \cite{Reisenegger92}.

Fig.~\ref{f:delta_eps}(b) shows that
for massive pure NSs the positive contribution in the inner core
%to the integral in Eq.~(\ref{e:pgw})
is dominant,
and thus the damping time in Fig.~\ref{f:tau-m}(a)
is about $10^6$--$10^8\,$s ($\sim$ days),
and decreases with the NS mass.
This means that the $g_1$ mode of a pure NS is likely
to be a stable and long-lasting source of GWs,
if no other very strong damping mechanisms dominate.
But for the hybrid stars,
the positive contributions of $\deps$ in the inner core are much larger,
and the damping time decreases very quickly by several orders of magnitude
with respect to pure NSs.
%Considering Eq.~(\ref{e:pgw}),
%it is mainly due to the high BV frequencies that can be reached in HSs.
Thus the behavior of the $g_1$-mode GW damping time
shows again a significant difference between HSs and pure NSs,
just like the frequencies.

%As for the $f$ and $p$ modes,
%their damping time is much smaller than for the $g$ modes.
%For a $2\ms$ NS and HS,
%the $f$-mode frequencies and damping times are quite similar.

The much smaller damping time of HSs indicates a much stronger GW strain
$|h_+|$, Eq.~(\ref{e:hp}).
This quantity depends on the oscillation amplitudes $W$ and $V$.
Their normalization can be determined from the total energy $E$ of oscillation
through Eq.~(\ref{e:osenergy}).
Choosing a typical energy scale $E \sim 10^{51}\,$erg \cite{Lugones21}
and a typical distance $D \sim 15\,$Mpc (star in the Virgo cluster),
we show in Fig.~\ref{f:tau-m}(b)
the GW strain amplitude of NS and HS $g_1$-mode oscillations.
One can see that $|h_+|$ for pure NSs with low masses is much lower,
with even a zero point,
which corresponds to the high damping time and its divergence in the upper panel.
For NSs with larger masses, $|h_+|$ increases with the NS mass,
while for HSs the equivalent irregular behavior as for $\taugw$ is exhibited.
Therefore,
our results for both the damping time and the strain amplitude
suggest stronger GW $g_1$-mode radiation of HSs than pure NSs,
which could thus be good observables to distinguish them from each other.

%------------------------------------------------------------------------------
\subsection{Prospects of observation}

Some of these features are likely to be detected
by the next generation of GW detectors
\cite{Regimbau16},
while the frequencies of $f$ and $p$ modes are not in the range of sensitivity
of current ground-based detectors.
Although the GW strain emitted by NSs in the Virgo cluster is only
of the order $10^{-25}$,
a $|h_+|$ of the order $10^{-22}$
could be obtained for NSs in our galaxy ($D\sim10\,$kpc),
which is within the detection ability of present GW detectors.
The minimum energy that should be released
to be detectable in present and planned GW observatories
can be estimated as \cite{Kokkotas01,Andersson11}:
\be
 \frac{E_\text{GW}}{\ms} = 3.5 \times 10^{36}
 \frac{1+4 Q^2}{4 Q^2}
 \frac{S_n}{1\text{s}}
 \left( \frac{S}{N} \frac{D}{10\text{kpc}} \frac{f}{1\text{kHz}} \right)^2
%\nonumber\\& \times
% \left(\frac{D}{10 \mathrm{kpc}}\right)^2
% \left(\frac{f}{1 \text{kHz}}\right)^2
% \left(\frac{S}{N}\right)^2
% \left(\frac{S_n}{1\text{s}} \right)
\:,
\label{e:egw}
\ee
where
$Q =\pi f\taugw$ is the quality factor,
$S_n$ is the noise power spectral density of the detector,
%$D$ is the distance to the source,
%$f$ is the frequency, $\taugw$ is the damping time,
and
$S/N$ is the signal-to-noise ratio.

Table~\ref{t:E_GW} lists some representative values,
for typical distances
$D=10\,$kpc (star in our galaxy)
and $D=15\,$Mpc (star in the Virgo cluster),
taking $S/N = 8$
and
$S_n^{1/2}=2\times10^{-23}\text{s}^{1/2}$
(representative of Advanced LIGO-Virgo at $\sim$ kHz frequency \cite{Abbott17a}),
and $S_n^{1/2}=10^{-24}\text{s}^{1/2}$
(illustrative of the planned third-generation ground-based
Einstein Observatory at the same frequencies \cite{Abbott16b}).
Although the minimum detectable energy of NSs in the Virgo cluster
is of the order $10^{52}$--$10^{53}\,$erg,
higher than the typical energy that can be released by a NRO $g$ mode,
the threshold for stars in our galaxy is only $10^{46}$--$10^{47}\,$erg,
much lower than this typical energy.
Therefore, those events could be detected by present and planned
GW observatories.

%------------------------------------------------------------------------------
\section{Conclusions}
\label{s:end}

In this work we investigated non-radial quadrupole oscillations
of cold and isolated NSs,
including pure NSs and HSs.
We adopted the BHF theory for NM, the DSM for QM,
and the Gibbs construction for their phase transition.
Based on the equilibrium structure,
we solved the equations for the non-radial $l=2$ oscillations
within Cowling approximation,
and obtained the radial and tangential displacement perturbations in NSs
as well as the eigenfrequencies of $g$, $f$, and $p$ modes for various EOSs.

The emergence of QM influences strongly the two kinds of sound speed
and the BV frequencies in HSs,
and consequently their $g$-mode oscillations.
We find eigenfrequencies $f_{g_1} \sim 300\,$Hz for pure NSs,
which increase very slowly with the NS mass,
while those of HSs increase very quickly with NS mass,
reaching above $700\,$Hz.
All these frequencies are in the sensitivity range
of current ground-based GW detectors.
This shows a clear difference of the $g$-mode frequency
between pure NSs and HSs,
which can thus be a good observable to distinguish them.
Such a difference is not obvious for the $f$ and $p$ modes.
The concurrent shorter $g_1$ damping times of HSs correspond
to larger GW strain and radiation power,
and thus easier detection than for pure NSs.
Estimates of the GW strain $h_+$ and minimum detectable energy $E_\text{GW}$
suggest that the GWs from the NRO $g_1$ mode of NSs/HSs in our galaxy
could be detected by present and planned detectors.
To sum up, the $g_1$ mode is the most suitable mode to
provide a window on the internal composition of the compact object.

In this work we disregarded the contribution of the crust
to the $g$-mode oscillations
by using the approximation $N_\text{crust}=0$.
In Ref.~\cite{Reisenegger92} the coupling between core and crust NRO modes
has been carefully analyzed and their mutual interference was found
sufficiently small, i.e.,
the frequency and damping time of the core modes are only weakly influenced
by the crust contribution,
especially for high-mass stars.
We have confirmed this conclusion by calculations
involving the Shen EOS $N_\text{crust}$,
which will be the subject of a separate paper.

The accurate computation of this feature requires in particular a careful
treatment of density discontinuities
due to changes of the chemical composition.
This might occur
inside the crust \cite{Finn87},
at the core-crust boundary,
or at the critical point from HM to QM
with a Maxwell phase transition \cite{Miniutti02,Rodriguez21,Lau20,Zhao22b}.
The associated discontinuity $g$ modes
might have similar frequencies as
and thus mix with the core modes \cite{Reisenegger92,Zhao22b}.

Apart from cold isolated NSs,
quadrupole oscillations also occur in various newly born NSs,
after NS mergers or supernova explosions,
which are expected to be more energetic and easier observable.
In such newly born NSs,
one needs to consider more realistic environment effects,
such as the EOS at finite temperature
\cite{Finn87,McDermott83},
the temperature/entropy distribution in NSs,
the neutrino trapping effects,
and also rotation.  %\cite{Panda16}.
We leave these to future work.

%\bigskip
\begin{acknowledgments}

We acknowledge financial support from the National Natural Science Foundation of China - Grant No. 12205260.

\end{acknowledgments}

%------------------------------------------------------------------------------
\newcommand{\epja}{Euro. Phys. J. A\ }
\newcommand{\aap}{Astron. Astrophys.\ }
\newcommand{\apjl}{Astrophys. J. Lett.\ }
\def\jcap{Journal of Cosmology and Astroparticle Physics}
\def\jcap{JCAP}
\newcommand{\mnras}{Mon. Not. R. Astron. Soc.\ }
\newcommand{\nphysa}{Nucl. Phys. A\ }
\newcommand{\physrep}{Phys. Rep.\ }
\newcommand{\plb}{Phys. Lett. B\ }
\newcommand{\ppnp}{Prog. Part. Nucl. Phys.\ }
\bibliographystyle{apsrev4-1}
\bibliography{nsnro}

\end{document}